\documentclass[useAMS,usenatbib]{mn2e}

\usepackage{psfig}
\usepackage{amsbsy}
\usepackage{aas_macros}
\newcommand{\msun}{\ensuremath{ {\rm M}_{\odot} }}

\newcommand\etal{et al.\ }

\newcommand{\captionfonts}{\footnotesize\bf}
\makeatletter  % Allow the use of @ in command names
\long\def\@makecaption#1#2{%
  \vskip\abovecaptionskip
  \sbox\@tempboxa{{\captionfonts #1: #2}}%
  \ifdim \wd\@tempboxa >\hsize
    {\captionfonts #1: #2\par}
  \else
    \hbox to\hsize{\hfil\box\@tempboxa\hfil}%
  \fi
  \vskip\belowcaptionskip}
\makeatother   % Cancel the effect of \makeatletter

\def\simlt{\lower.5ex\hbox{$\; \buildrel < \over \sim \;$}}
\def\simgt{\lower.5ex\hbox{$\; \buildrel > \over \sim \;$}}

\newcommand{\thetab}{\mbox{\boldmath $\theta$}}
\newcommand{\be}{\begin{equation}}
\newcommand{\ee}{\end{equation}}
\newcommand{\ba}{\begin{eqnarray}}
\newcommand{\ea}{\end{eqnarray}}
\newcommand{\de}{\partial}
\newcommand{\llb}{\mbox{\boldmath $\ell$}}
\newcommand{\lgl}{\langle}
\newcommand{\rgl}{\rangle}
\newcommand{\rb}{\mbox{\boldmath $r$}}

\title[SUNGLASS: A new weak  lensing simulation pipeline ]
{SUNGLASS: A  new weak lensing simulation pipeline}
\author[Kiessling     \etal]    {A.     Kiessling$^{1}$\thanks{E-mail:
    aak@roe.ac.uk},     A.       F.      Heavens$^{1}$,     A.      N.
  Taylor$^{1}$\\   $^{1}$Institute   for   Astronomy,  University   of
  Edinburgh, Royal Observatory, Blackford Hill, Edinburgh, U.K.}

\begin{document}

\date{Accepted --. Received --; in original form --.}

\pagerange{\pageref{firstpage}--\pageref{lastpage}} \pubyear{0000}

\maketitle

\label{firstpage}

\begin{abstract}
A  new  cosmic  shear  analysis pipeline  {\small  SUNGLASS}  ({\small
  S}imulated  {\small UN}iverses  for  {\small G}ravitational  {\small
  L}ensing {\small  A}nalysis and {\small S}hear  {\small S}urveys) is
introduced.  {\small  SUNGLASS} is  a pipeline that  rapidly generates
simulated universes  for weak lensing  and cosmic shear  analysis. The
pipeline forms suites of  cosmological N-body simulations and performs
tomographic  cosmic  shear  analysis using  line-of-sight  integration
through  these   simulations  while  saving   the  particle  lightcone
information. Galaxy shear and convergence catalogues with realistic 3D
galaxy redshift distributions are produced for the purposes of testing
weak  lensing analysis techniques  and generating  covariance matrices
for data analysis and cosmological parameter estimation.  We present a
suite of fast medium resolution simulations with shear and convergence
maps for a generic 100 square degree  survey out to a redshift of $z =
1.5$, with  angular power spectra  agreeing with the theory  to better
than  a few  percent  accuracy up  to  $\ell =  10^3$  for all  source
redshifts up to $z = 1.5$ and  wavenumbers up to $\ell = 2000$ for the
source  redshifts $z  \ge 1.1$.   At  higher wavenumbers,  there is  a
failure of the theoretical lensing power spectrum reflecting the known
discrepancy  of the  \cite{spj+03}  fitting formula  at high  physical
wavenumbers.    A  two-parameter   Gaussian  likelihood   analysis  of
$\sigma_8$  and   $\Omega_m$  is  also  performed  on   the  suite  of
simulations,  demonstrating  that   the  cosmological  parameters  are
recovered from the simulations  and the covariance matrices are stable
for  data analysis.   We find  no  significant bias  in the  parameter
estimation  at  the  level  of  $\sim 0.02$.   The  {\small  SUNGLASS}
pipeline should be an invaluable tool in weak lensing analysis.

\end{abstract}


\section{Introduction}
Cosmic  shear analysis  is an  excellent method  for probing  the dark
Universe       \citep[for      reviews,       see][and      references
  therein]{m99,bs01,r03a,sch05,mvv+08,mkr10}.  It is also a reasonably
new field of research with  cosmic shear first being observed just ten
years   ago  \citep{bre00,kwl00,wme+00,wtk+00}.    Weak  gravitational
lensing effects on  a cosmic scale are a mere 1\%  change in shape and
observational  systematics  makes  the  measurement of  these  changes
challenging.   However,  the   combination   of  the   well-understood
underlying  physics   and  the  expected   precision  of  cosmological
parameter estimation make the effort worthwhile.

Next generation  telescope surveys will  observe more of the  sky than
ever before and the volume of data they will produce is unprecedented.
Future  surveys promise  to determine  the equation  of state  of dark
energy  to  1\%  as  well   as  probing  the  possibilities  of  extra
dimensional  gravity models  and alternative  cosmologies.   The first
Pan-STARRS\footnote{Pan-STARRS       http://pan-starrs.ifa.hawaii.edu/}
telescope is currently undertaking a cosmic shear survey of the entire
visible  sky from  its location  in Hawaii  and new  projects  such as
VST-KIDS\footnote{VST-KIDS   http://www.astro-wise.org/projects/KIDS/},
DES\footnote{DES                    https://www.darkenergysurvey.org/},
HALO\footnote{Rhodes  \etal,  in  preparation}, Euclid\footnote{Euclid
  http://sci.esa.int/euclid}           and          LSST\footnote{LSST
  http://www.lsst.org/} and  are planned to perform  wide field cosmic
shear  surveys,  measuring  both  large,  linear  scales,  and  small,
non-linear scales.

Due to  the relative youth of  this field, techniques  are still being
developed  to exploit  the weak  lensing  data from  these surveys  to
provide  further understanding  on  the nature  of  the Universe.   To
realise the potential  of these new telescope surveys  and to test new
weak lensing analysis techniques,  challenges must be met.  To achieve
the  small  statistical  errors  required,  experiments  require  full
end-to-end simulations of huge volumes which also probe the non-linear
regime  to assist  in understanding  the limitations  of  the analysis
techniques.  Simulations  offer data sets with  known parameters which
are essential  when testing analysis pipelines.   Simulations can also
include effects which may be difficult to model theoretically, such as
source clustering and galaxy  alignments, as well as other systematics
and  real-world effects.   An additional  role for  simulations  is in
accurate estimation  of the covariance of  observable quantities. This
is needed for the analysis  of surveys and analytic approximations can
be wholly inadequate \citep[e.g.][]{svh+07}.  Monte Carlo analysis can
be performed with simulations  to provide covariance matrices that are
required  for  data analysis  and  cosmological parameter  estimation.
Simulations are also required  for rigorous testing and development so
all  analysis  methods  can   be  analysed  blindly  before  the  same
techniques are applied to real data.  To address these challenges, the
{\small SUNGLASS},  {\small S}imulated {\small  UN}iverses for {\small
  G}ravitational {\small L}ensing {\small A}nalysis and {\small S}hear
{\small S}urveys,  pipeline has been developed  to produce simulations
and mock shear and convergence catalogues rapidly for weak lensing and
cosmic  shear analysis.   The purpose  of this  paper is  to introduce
{\small SUNGLASS} and show rigorous testing of its outputs.

Many weak lensing studies use simulations with very high resolution to
run  their analysis  \citep[e.g.][]{fgc+08,hhw+09,tpp+09,shj+10}.  The
computational  cost   of  running   these  simulations  is   high  and
consequently  there  is often  only  a  single realisation  available.
However,  it is  very  important to  ensure  that covariance  matrices
calculated from these simulations are not contaminated by correlations
in  the simulations  \citep{hss07}.  In  order to  ensure uncorrelated
data, a Monte  Carlo suite of simulations should  be used to determine
the covariance  matrix \citep{sht+09}.  In this work,  100 independent
simulations were constructed using {\small SUNGLASS}.

To  date, there  are  still reasonably  few  weak lensing  simulations
available.   Of   the  few  that  are  available,   many  implement  a
ray-tracing technique where light rays are propagated from an observer
to            a             lensing            source            plane
\citep[e.g.][]{jsw00,vw03,hhw+09,tpp+09,sht+09,dh10,vlv+10}.
Ray-tracing  is  computationally  intensive  and time  consuming  when
solving the  full ray-tracing equations. If the  Born approximation is
used in the  ray-tracing, the time to run the  analysis is reduced but
the process is still computationally intensive and the simulation data
still  needs  to  be  binned   in  three  dimensions  to  perform  the
calculations.   An  alternative   to   ray-tracing  is   line-of-sight
integration, which  uses the  Born approximation to  calculate rapidly
the      weak     lensing      signal     through      a     lightcone
\citep[e.g.][]{wh00,fgc+08}. This method is not suitable in the strong
lensing  regime  but in  the  weak lensing  regime,  it  is rapid  and
requires  fewer computational  resources than  ray-tracing techniques.
In this paper, a  new line-of-sight integration technique, implemented
in the  {\small SUNGLASS} pipeline,  for measuring convergences  in an
N-body simulation is  introduced. This new method is  rapid and can be
run  on a  single  processor of  a  desktop computer.  In contrast  to
ray-tracing, the  method does not  bin in the radial  direction, using
all of the redshift information available. Although the catalogues are
suitable for real-space analysis, {\small SUNGLASS} analyses and tests
our mock weak  lensing surveys in Fourier space,  using power spectra,
as it is possible to  cleanly distinguish between linear and nonlinear
regimes in Fourier space.  We  are also able to easily identify scales
where the  simulations are reliable  by determining the region  of the
power  spectrum in Fourier  space that  lies between  the size  of the
simulated  volume at low  wavenumbers and  shot-noise due  to particle
discreteness and pixelization effects at high wavenumbers.

The outline  of this paper is as  follows.  Section \ref{sec:sunglass}
introduces the {\small SUNGLASS} pipeline.  Details of the simulations
are in section \ref{sec:sims} and the line-of-sight integration method
for  determining  shear  and  convergence without  radial  binning  is
described in section  \ref{sec:map}. Section \ref{sec:ps} presents the
shear   and   convergence   power   spectrum  analysis   and   section
\ref{sec:mock}  deals with  the generation  of the  mock  galaxy shear
catalogues.   An application of  the mock  catalogues is  discussed in
section   \ref{sec:Like}  where   Gaussian  likelihood   estimates  of
$\Omega_m$ and $\sigma_8$ are performed. A summary of the pipeline and
methods concludes the paper in section \ref{sec:disc}.

\section{Details of the SUNGLASS pipeline}
\label{sec:sunglass}

{\small  SUNGLASS}  is a  pipeline  that  generates  cosmic shear  and
convergence catalogues using  N-body simulations. The pipeline creates
mock galaxy shear catalogues that can be used to test the cosmic shear
analysis software used  on telescope survey data sets.   The nature of
the pipeline also allows  many simulation realisations to be generated
rapidly  to   produce  covariance  matrices  for   data  analysis  and
cosmological parameter  estimation. The pipeline begins  by creating a
suite  of cosmological N-body  simulations.  Lightcones  are generated
through the simulations and tomographic shear and convergence maps are
determined using line-of-sight integrations at multiple lensing source
redshifts.  Finally,  mock galaxy catalogues  with fully 3D  shear and
convergence   information  and   galaxy  redshift   distributions  are
assembled  from   the  lightcones   and  the  tomographic   shear  and
convergence planes.   The following sections  detail each step  of the
{\small SUNGLASS} pipeline.

\subsection{The N-body Simulations}
\label{sec:sims}
All of  the simulations presented  in this work  were run on  a modest
Xeon cluster,  using 4 nodes with  dual Xeon E5520  2.27 GHz quad-core
processors per node and 24Gb  shared memory per node.  The simulations
were run  using the cosmological structure  formation software package
{\small GADGET2} \citep{sp05}. {\small GADGET2} represents bodies by a
large  number, \textit{N} (in  this work  we use  $512^3$), particles.
Each particle is  `tagged' with its own unique  kinematic and physical
properties that evolve with  the particle over time.  {\small GADGET2}
models the  dynamics of dark  matter particles using a  Tree-PM scheme
and for  the purposes  of this work,  only dark matter  particles were
considered.

The pre-initial particle distribution for the simulations used in this
work  is a  \textit{glass} which  has sub-Poissonian  noise properties
\citep{w94}.  This distribution has no preferred direction with forces
on  each particle being  close to  zero.  If  a glass  is used  as the
initial condition in a  standard integrator, structures do not evolve.
Particle  displacements are  imposed manually  as an  initial  step to
enable structure formation.  The initial power spectrum was imposed on
the particles using the  parallel initial conditions generator {\small
  N-GenIC} that was provided  by Volker Springel. The initial particle
displacement  field is  formed by  using the  Zel'dovich approximation
\citep{z70} to  perturb the particles, imposing  an \cite{eh98} matter
power spectrum on  the particles, and giving each  particle an initial
velocity.

Multiple   medium-resolution  simulations   were   run  with   $512^3$
dark-matter particles, in  a box of $L =  512h^{-1}$~Mpc comoving side
length with periodic  boundary conditions.  The following cosmological
parameters  were  used  for  a  flat  concordance  $\Lambda$CDM  model
consistent    with   the    WMAP   7-year    results   \citep{jbd+10}:
$\Omega_{\Lambda} = 0.73$, $\Omega_m = 0.27$, $\Omega_b = 0.045$, $n_s
= 0.96$, $\sigma_8 = 0.8$ and $h = 0.71$ in units of 100 km~ s$^{-1}$~
Mpc$^{-1}$.  The  particle mass is  $7.5 \times 10^{10}$\msun  and the
softening length  is $33h^{-1}$~kpc. The simulations  were all started
from a redshift of $z = 60$ and allowed to evolve to the present.

The simulation data were stored  at 26 output times corresponding to a
$128h^{-1}$~Mpc  comoving  separation,  between  $z  =  1.5$  and  the
present.  These  snapshots were chosen to fall  within the photometric
redshift  error   of  $\sigma_z   <  0.05(1+z)$  corresponding   to  a
displacement  of  $  \simeq  147h^{-1}$~Mpc  at $z=1$.  In  a  $512^3$
particle  simulation, this amounts  to 100GB  data per  simulation and
takes approximately 21hrs to run on the Xeon cluster's 32 processors.

\subsection{Shear and Convergence Map Generation}
\label{sec:map}

We begin by  determining the shear and convergence  for a source plane
at  fixed comoving  distance  $r_s$.  We  consider  a distribution  of
sources in Section \ref{sec:mock}.

The effects of weak gravitational lensing on a source can be described
by  two  fields,  the  spin-2  shear, $\gamma$,  which  describes  the
stretching  or compression  of  an image,  and  a scalar  convergence,
$\kappa$, which  describes its change  in angular size.  These  can be
related to a lensing potential field, $\phi$, by
\begin{eqnarray} 
  \kappa &=& \frac{1}{2}\de^2 \phi,\\ 
  \gamma &=& \gamma_1 + i \gamma_2 = \frac{1}{2}\de \de \phi,
\end{eqnarray}
where $\gamma_1$  and $\gamma_2$ are the orthogonal  components of the
shear distortion, and  $\de= \de_x + i \de_y$  is a complex derivative
on the sky.

We  want to  generate shear  and  convergence maps  along a  lightcone
through the simulation.  Instead of using ray tracing to determine the
lightcone  \citep[e.g.][]{wco98,jsw00,tpp+09,hhw+09},  a line-of-sight
integration  was implemented  using the  Born approximation  where one
integrates   along  an  unperturbed   path  \citep[e.g.][]{ch02,vw03}.
\cite{fgc+08} build their convergence maps by adding slices from their
simulation with  the appropriate lensing  weight and averaging  over a
pixel;
 \begin{eqnarray}
  \bar\kappa(\thetab_i,r_s)                                         &=&
  \int_0^{r_s}dr~K(r,r_s)~\bar\delta(\thetab_i,r)  \\ &=& \frac{3H_0^2
    \Omega_m}{2c^2}   \sum_j  ~  \bar\delta(\thetab_i,r_j)   ~  \frac{
    (r_s-r_j) r_j} {r_sa_j} ~ \Delta r_j, \label{eq:kappa}
 \end{eqnarray}
where $\mathbf{\theta}_i$ is the position if the $i^{th}$ pixel on the
sky and $j$ is a bin in the radial direction which is at a distance of
$r_j$ and has a width of  $\Delta r_j$. An overline denotes an average
over a pixel  on the sky. The expansion factor at  each radial bin $j$
is  given by $a_j$  and the  comoving radial  distance of  the lensing
source plane is given by  $r_s$.  In order to make these calculations,
the 3D  matter overdensity $\bar\delta(\thetab,r)$  must be calculated
by binning the simulation data in three dimensions.

A limitation of  this approach is memory, speed  and accuracy. Here we
propose,  in the  {\small SUNGLASS}  pipeline,  a new  method for  the
line-of-sight  integration so that  no radial  binning is  required to
determine the convergence.  The particles are binned in a fine angular
grid while allowing them to keep their radial co-ordinate.

Rewriting equation (\ref{eq:kappa}) we find the average convergence in
an angular pixel, with no radial binning, is given by
 \begin{equation}
  \bar\kappa_{p}(r_s)      =       \sum_j      \frac{K      (r_j,r_s)}
            {\Delta\Omega_p\bar{n}  (r_j)  r_j^2}  -  \int_0^{r_s}  \!
            dr~K(r,r_s),
 \label{eq:convergence}
 \end{equation}
where  $\Delta\Omega_p =  \Delta\theta_x\Delta\theta_y$  is the  pixel
area and $K(r,r_s)$ is the scaled lensing kernel:
 \begin{equation}
    K(r,r_s) = \frac{3H_0^2\Omega_m}{2c^2}\frac{(r_s-r)r}{r_sa(r)}.
 \end{equation}
Hereafter we drop the overline and assume all fields are averaged over
an angular  pixel. A derivation of  equation (\ref{eq:convergence}) is
given    in   Appendix    \ref{ap:conv}.     In   practice    equation
(\ref{eq:convergence})  can be  calculated by  a running  summation so
that it is not necessary  to re-calculate the convergence from scratch
for each source redshift.

\begin{figure}
  \centerline{\psfig{file=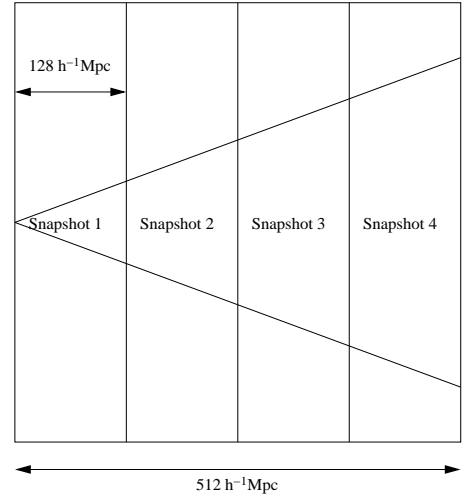,width=0.7\linewidth,clip=}}
  \caption{Lightcone  geometry through  a simulation  box  volume. The
    lightcone travels  through the first $128h^{-1}~$Mpc  of the first
    simulation  and   then  the  next  $128h^{-1}~$Mpc   of  the  next
    simulation  etc. At  the end  of the  simulation volume,  the next
    volume  snapshots have  their centroids  shifted and  are randomly
    rotated to avoid repeated structures along the lightcone.}
  \label{fig:lightcone}
\end{figure}

The convergence maps  are generated by adding the  particles that fall
within  the  lightcone  to  the line-of-sight  integration.   To  show
evolution through the lightcone, the simulation volumes are split into
128$h^{-1}$~Mpc sections.  The first  128$h^{-1}$~Mpc of the first ($z
= 0$) snapshot is used, the second 128$h^{-1}$~Mpc of the second ($z >
0$) snapshot and  so on until the end of the  simulation box volume is
reached  at snapshot 4  as shown  in Figure  \ref{fig:lightcone}.  The
centroid of the next simulation box is then shifted and the simulation
box is rotated randomly to  try to avoid repeated structures along the
line-of-sight \citep[e.g.][]{wh00,vw03}. The boxes are always periodic
in  the  transverse  direction.  This  continues through  all  of  the
snapshots out  to a  redshift of $z=1.5$.   The source  redshifts have
been  placed  at  $\Delta  z=0.1$  intervals  because  the  change  in
convergence  between  these redshifts  is  small  enough that  desired
redshift   values  in   between  can   be  accurately   determined  by
interpolation.

Once  the convergences  have been  calculated  at each  of the  source
redshifts,  the shear  values can  be  determined on  a flat-sky.  The
flat-sky shear and convergence Fourier coefficients are related by
\begin{eqnarray}
  \gamma_1(\boldsymbol{\ell})       &=&      \kappa(\boldsymbol{\ell})
  \frac{(\ell_x^2-\ell_y^2)} {(\ell_x^2+\ell_y^2)},\\
  \gamma_2(\boldsymbol{\ell})       &=&      \kappa(\boldsymbol{\ell})
  \frac{2\ell_x\ell_y} {l_x^2+l_y^2},
  \label{shearequ}
\end{eqnarray}
where  $\kappa(\boldsymbol{\ell})$  is the  Fourier  transform of  the
convergence and $\ell_x$ and  $\ell_y$ are the Fourier variables.  The
Fast Fourier transform used throughout this paper is FFTW\footnote{The
  Fastest  Fourier Transform  in the  West  http://www.fftw.org/}. The
periodic nature  of FFTW  requires that the  field is buffered  with a
small number  of bins that  are trimmed off  after the shear  has been
calculated.  To  test  the  algorithm  we also  estimated  B-modes  by
calculating the unphysical imaginary  part of the convergence $\beta =
\mathrm{imag}(\kappa)$, from the shear,
\begin{eqnarray}
  \beta(\llb)        =        \gamma_1(\llb)\left(\frac{2\ell_x\ell_y}
       {l_x^2+l_y^2}\right)                             +\gamma_2(\llb)
       \left(\frac{\ell_x^2-\ell_y^2} {\ell_x^2+\ell_y^2}\right).
\end{eqnarray}

\begin{figure}
 \psfig{file=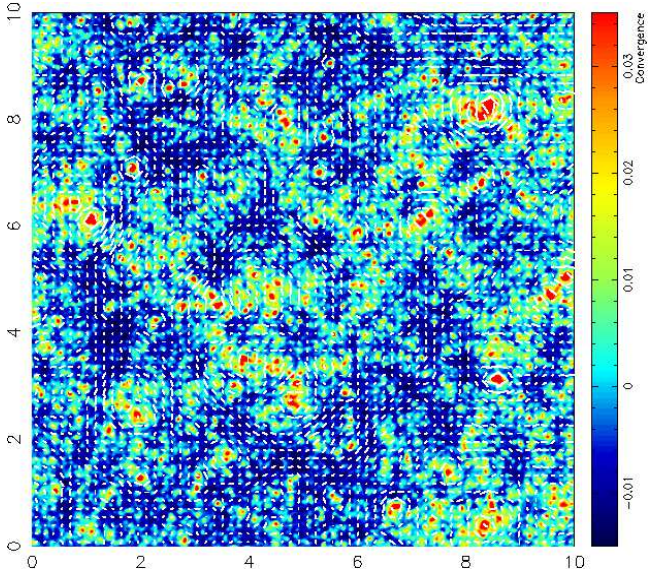,angle=-90,width=\linewidth,clip=}
 \caption{Convergence  and shear  map for  a simulated  survey  of 100
   square  degrees with  a  single source  redshift  of $z=0.8$.   The
   colour-scale background shows the convergence while the white ticks
   show the shear signal.} 
\label{fig:map}
\end{figure}

Figure \ref{fig:map} is an example  of a convergence and shear map for
a  field that  is 100  square degrees  at a  source redshift  of  $z =
0.8$. There are 2048 bins  in each transverse direction and no binning
in  the  radial  direction.   The  background of  the  map  shows  the
integrated convergence  along the  lightcone up to  $z = 0.8$  and the
white ticks show the shear at  this source redshift. The length of the
ticks  has been  multiplied  by  an arbitrary  constant  to make  them
visible as  the magnitude of the  shear is at the  percent level.  The
red patches show areas of  the highest convergence and the shear ticks
clearly trace these regions  tangentially. These maps can be generated
for  the  standard  simulations  at multiple  source  redshifts  quite
rapidly once the  simulations have been run.  The  most time consuming
module  in  this  code  is  reading  in the  snapshots  due  to  their
reasonably large size of 100GB.  This module can be optimised by using
the  fastest available  data transfer  rates  on the  drive where  the
snapshot data is stored.

\subsection{Shear and Convergence Power Spectra}
\label{sec:ps}

\begin{figure*}
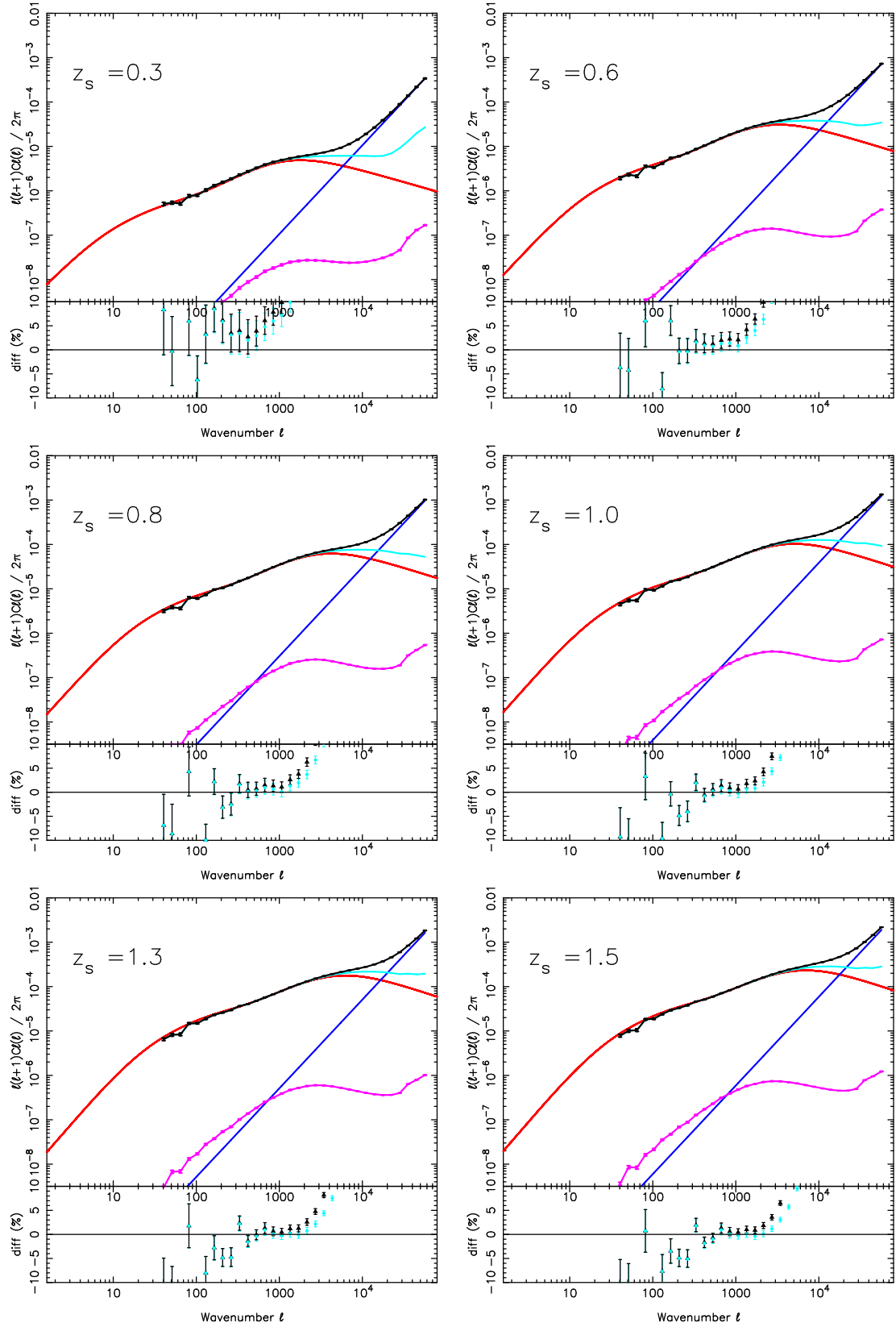

\begin{minipage}{148mm}
  \begin{tabular}{cc}
    \psfig{file=PSwErrors_gamma_03.ps,angle=-90,width=0.475\linewidth,clip=} &
    \psfig{file=PSwErrors_gamma_06.ps,angle=-90,width=0.475\linewidth,clip=}\\

    \psfig{file=PSwErrors_gamma_08.ps,angle=-90,width=0.475\linewidth,clip=} &
    \psfig{file=PSwErrors_gamma_10.ps,angle=-90,width=0.475\linewidth,clip=}\\

    \psfig{file=PSwErrors_gamma_13.ps,angle=-90,width=0.475\linewidth,clip=} &
    \psfig{file=PSwErrors_gamma_15.ps,angle=-90,width=0.475\linewidth,clip=}

  \end{tabular}
\end{minipage}
\caption{Simulated  slices  of  the  shear power  spectra  for  N-body
  particle data  at source redshifts of $z=0.3$,  $0.6$, $0.8$, $1.0$,
  $1.3$  and  $1.5$.  The  smooth  (red)  line  shows the  theoretical
  predictions,  the straight  diagonal  (blue) line  is the  predicted
  shot-noise at each  source redshift.  The black points  are the mean
  power spectrum of  the simulated data for the  100 realisations with
  errors  on the mean  shown and  the (light  blue) curve  under these
  points is  the simulation data with the  shot-noise subtracted.  The
  sub-shot-noise (magenta) curve is  the estimated induced B-mode. The
  lower panel  shows the fractional percentage  difference between the
  simulated  shear power  and  the theoretical  prediction with  black
  points  representing  the  simulated  data  and  light  blue  points
  representing the shot-noise subtracted simulation data.}
\label{fig:PSComp}
\end{figure*}

In order to verify the accuracy of the shear and convergence maps, the
shear  and convergence power  spectra are  determined for  each source
redshift. From  equation (\ref{eq:kappa}), the  theoretical prediction
for the shear  and convergence power spectrum for  sources at redshift
$z$ is given by
 \begin{equation}
    \label{shearpower}
    C_\ell^{\gamma\gamma}(z)  =   C_\ell^{\kappa\kappa}(z)  =  \frac{9
      H_0^4    \Omega_m^2}{4    c^4}    \int_0^{r_s}    \!    dr    \,
    P\left(\frac{\ell}{r};r\right)  \frac  {[r_s(z)  -  r]^2}{r_s^2(z)
      a^2(r)},
 \end{equation}
\citep{mvv+08}  where $P(\ell/r;r)$  is  the 3D  matter density  power
spectrum at a redshift $z$.

From  the simulations it  is possible  to determine  an angle-averaged
power  spectrum  from the  convergence  and  shear  calculated in  the
lightcones. When  taking in to  consideration the conventions  used in
FFTW,  the  discretised convergence  power  spectrum  for  a slice  in
redshift is given  as the sum over logarithmic  shells in $\ell$-space
as
 \begin{equation}
  \frac{\ell(\ell+1)   \hat{   C}_\ell^{\kappa   \kappa}(z)}{2\pi}   =
  \sum_{\ell  {\rm \,  in \,  shell}}  \frac{|\kappa(\llb,z)|^2} {n^2~
    \Delta \ln \ell},
 \end{equation}
where $n$  is the total  number of bins  in the Fourier  transform and
$\Delta  \ln  \ell$ represents  the  thickness  of  the shell  in  log
$\ell$-space,  and   $\hat{C}_\ell^{\kappa\kappa}$  is  the  estimated
power. Similarly the shear power is estimated by
 \begin{equation}
  \frac{\ell(\ell+1)    \hat{C}_\ell^{\gamma    \gamma}(z)}{2\pi}    =
  \sum_{\ell  {\rm   \,  in  \,   shell}}\frac{|\gamma_1(\llb,z)|^2  +
    |\gamma_2(\llb,z)|^2} {n^2~ \Delta \ln \ell}.
 \end{equation}
The B-mode power is estimated in the same way as the convergence.

The modes in this power spectrum  are arranged on a square grid, which
causes discreteness errors when binned  in annuli at small $\ell$.  To
correct for  this, the power  is scaled by  the ratio of  the measured
number of modes to the expected number of modes,
 \begin{equation}
    N_{exp} = g \pi (\ell_{max}^2 - \ell_{min}^2),
 \end{equation}
where $g  = (L/2\pi)^2$ is the density  of states, $L$ is  the size of
the field  in radians, $\ell_{max}$  and $\ell_{min}$ are  the minimum
and  maximum  wave  numbers  in   this  shell.   The  effect  of  this
normalisation  correction is about  $10\%$ at  the lower  wave numbers
while   the  higher   wavenumbers  remain   largely   unaffected.  The
discreteness correction  is not perfect  which is why the  same slight
zig-zag of the power spectrum is evident in all of the source redshift
planes at wavenumbers $\ell < 100$.

We can compare our simulated  shear and convergence power spectra with
the theoretical expectation. The  theoretical power spectrum we use is
determined using a code kindly provided by Benjamin Joachimi \citep[as
  demonstrated      in][and       extensively      tested      against
  iCosmo\footnote{http://www.icosmo.org}\citep{rak+08}]{js08,js09,js10}.
This code  uses the method  of \cite{spj+03} for the  non-linear power
spectrum,  the  matter  transfer   function  of  \cite{eh98}  and  the
analytical  expression  for  the  linear  growth factor  as  given  in
\cite{h77}.

Due to the  discrete number of particles in  an N-body simulation, the
measured power spectrum  measured will be the combined  real shear and
convergence power plus a shot-noise power contribution,
\begin{eqnarray}
  \hat{C}^{\kappa\kappa}_\ell      =     C^{\kappa\kappa}_{\ell}     +
  C_{\ell}^{SN},
\end{eqnarray}
where  $\hat C^{\kappa\kappa}_\ell$  is the  power estimated  from the
simulation.   The  shot-noise  power  can  be  derived  from  equation
(\ref{shearpower}) using a  white-noise power spectrum, $P_{SN}(k,r) =
1/\bar{n}_3(r)$, where $\bar{n}_3(r)$ is  the 3-D mean comoving number
density of particles  in the simulation. The shot-noise  power for the
shear and convergence is then given by
 \begin{equation}
    C_{\ell}^{SN} = \frac{9 H_0^4 \Omega_m^2}{4 c^4}\int_0^{r_s} \! dr
    \, \frac{(r_s-r)^2}{\bar{n}(r) r_s^2a(r)^2}.
 \end{equation}
Usually,  for simulated  particles, $\bar{n}$  will be  a  constant in
comoving coordinates.

Figure {\ref{fig:PSComp}}  shows the mean, normalised  2-D shear power
spectra estimated  from 100 independent simulations  (black points and
line),  with the  error  bars  showing the  scatter  on the  estimated
mean. The figures show the shear  power for sources at redshifts of $z
= 0.3,~ 0.6,  ~ 0.8,~ 1.0,~ 1.3~\textrm{and}~ 1.5$.   The smooth (red)
line shows the theoretical  prediction for the ensemble-averaged shear
power spectrum,  while the diagonal  (blue) lines show  the shot-noise
power for  each source redshift.   The (light blue) curve  between the
simulated data and the theory curve shows the mean power spectrum with
the expected shot-noise subtracted and the lower (magenta) curve shows
the estimated B-mode power spectrum.

The  bottom  panel of  each  figure  shows  the percentage  difference
between  the measured  shear power  spectrum and  the ensemble-average
theory prediction  (black), while the  lower (light blue)  points show
the  shot-noise subtracted  shear  power spectrum.   Overall the  mean
shear  power  agrees   well,  to  within  a  few   percent,  with  the
ensemble-averaged  theoretical  model over  the  $\ell$-range $\ell  <
1000$ for all source redshifts. The difference of a few percent is due
to the fact that the theory  3D matter density power spectrum is a few
percent lower  than the measured data power  spectrum. Calculating the
highly non-linear  power spectrum is  currently not accurate to  a few
percent and many  calculations of this theory curve  do not agree with
each other to within a few  percent. The Joachimi theory curve was the
closest  fit  to the  simulations  and  was  used for  all  subsequent
calculations.  At low $\ell$ the measured signal drops as we reach the
size of the  simulation box, while at high  $\ell$, the estimated mean
shear power  becomes shot-noise dominated before  reaching the highest
mode  allowed   by  the  resolution  of  the   angular  pixels  beyond
$\ell=1/\theta_{\rm pix } \simeq 10^4$.

Before  reaching   pixel-resolution,  the  measured   shear  power  at
high-$\ell$ agrees well with  the predicted shot-noise. This agreement
suggests that  the shot-noise  model works well  in this  regime, even
though the initial particle distribution is a glass \citep[see] [for a
  discussion]{bge95}. This  suggests an improved estimate  of the mean
power  can be found  by subtracting  off the  shot-noise contribution.
However,  the shot-noise subtracted  shear power  does not  follow the
ensemble-averaged  theoretical power estimated  from the  theory code.
It is likely this is a  failure of the theoretical model of lensing --
on  small-scales  the \cite{spj+03}  nonlinear  correction formula  is
known to  underestimate the matter-density power  spectrum, $P(k)$, by
up to 10\% at wavenumbers of $k <  1$ and as great as 50\% at $k = 10$
Mpc$^{-1}$ (Giocoli,  private communication) and hence  has been shown
to underestimate  the shear  and convergence power  spectrum by  up to
30\% on  scales of  $\ell < 10^4$  \citep{hhw+09}.  In the  absence of
accurate fitting  formulae, simulations  like those presented  in this
paper may  be used to improve theoretical  predictions.  However, this
needs to be  explored in more detail before it  is fully understood so
in subsequent analysis in this  paper we will restrict our analysis to
the  region  of the  measured  power  spectrum  that agrees  with  the
theoretical prediction.

Figure  {\ref{fig:PSComp}}  also  shows  the  estimated  B-mode  power
spectra. When  galaxies trace the  shear signal, we expect  the B-mode
power to pick up a shot-noise dependence. But here the shear signal is
a pixelized field  which would be continuous in  the limit of infinite
pixels.  Therefore we  do not  expect  there to  be a  noise-generated
B-mode.  However, B-modes  can still  be generated  due to  leakage of
power from the convergence field  caused by the finite window function
when we generate the shear  field from equation (\ref{shearequ}). As a
consequence the induced  B-mode has the shape of  the shear power, but
suppressed by around three orders of magnitude.

In this section we have shown that the {\small SUNGLASS} algorithm for
calculating the  shear and convergence  maps and the power  spectra in
redshift slices  is accurate  to a  few percent over  a wide  range of
scales and redshifts.   Wavenumbers up to $1500$ can  be recovered for
the source redshifts $z \ge 1.1$ with this simulation resolution.  For
shot-noise  subtracted  power spectra,  the  recovered modes  increase
before the angular pixel resolution cuts off the power.

\subsection{Mock 3-D Weak Lensing Galaxy Catalogues}
\label{sec:mock}

\begin{table}
 \begin{center}
    \begin{tabular}{cccccc}
      \hline\\
         $N_{\rm Surveys}$ & Area & $n$  & $z_{\rm median}$ &
         $\sigma_0$  & $z_{\rm max}$ \\
         \hline \\
        100 & 100   &  15 & 0.82        & 0.05   & 1.5 \\
        \hline
    \end{tabular}
 \end{center}
 \caption{Table of  mock weak lensing  survey parameters used  in this
   paper.}
 \label{surveytable}
\end{table}

Real, 3-D weak lensing data  analysis is applied to a galaxy catalogue
where  galaxy angular positions  and redshift  are added  to estimated
shears for each  galaxy. For a 2-D analysis,  individual redshifts are
ignored and  the theory  uses only the  redshift distribution.   It is
straightforward  to generate  a simple  3-D mock  weak  lensing galaxy
catalogue  with the information  in the  lightcones we  have generated
from  the simulations. Shear  and convergence  maps are  generated for
each lensing source redshift and  then each particle in the simulation
is assigned a shear  and convergence by interpolating between adjacent
planes. The  error introduced by linearly interpolating  the shear and
convergence between source redshift planes separated by $\Delta z=0.1$
was estimated  by comparing  with much higher  redshift-sampled planes
and  found  to  be  substantially  below  the  theoretical  prediction
($\Delta   C_{\ell}^{\gamma   \gamma}<10^{-7}$)   except  at   angular
wavenumbers where shot-noise  becomes dominant.  With the interpolated
shear  and  convergence assigned  to  each  particle,  we now  have  a
fully-sampled  3-D mock weak  lensing galaxy  catalogue, which  can be
down-sampled to generate realistic weak lensing surveys.

\begin{figure}
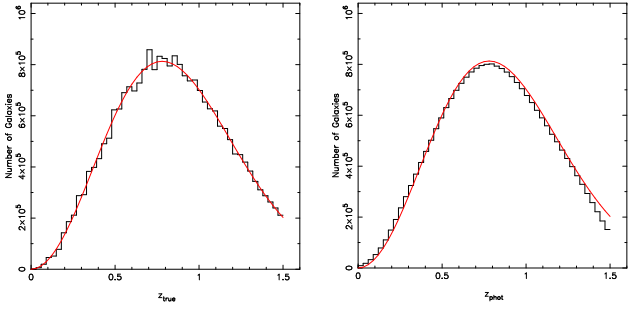

  \centering
  \begin{tabular}{cc}

    \psfig{file=dndz.ps,angle=-90,width=0.46\linewidth,clip=} &
    \psfig{file=dndzphot.ps,angle=-90,width=0.46\linewidth,clip=}

  \end{tabular}
  \caption{\textit{Left:} The galaxy distribution, $n(z)$, in the mock
    galaxy  catalogue. The  smooth  (red) line  shows the  theoretical
    $n(z)$  and the  black  histogram shows  the  distribution from  a
    single  simulation lightcone.  The  histogram shows  the clustered
    nature of the  lightcone.  \textit{Right:} The galaxy distribution
    in  the mock  galaxy  catalogue with  photometric redshift  errors
    assigned  to each  galaxy.   The structures  visible  in the  true
    redshift lightcone have been smoothed out with the addition of the
    photo-z errors.}
  \label{fig:dndz}
\end{figure}

To  down-sample  the full  3-D  weak  lensing  simulated lightcone  to
construct  a realistic  3-D weak  lensing galaxy  catalogue, we  use a
galaxy redshift distribution \citep{rmr+04}
 \begin{equation}
    n(z)      \propto     z^{\alpha}      \exp\left[      -     \left(
      \frac{z}{z_0}\right)^{\beta}\right],
    \label{eq:nz}
 \end{equation}
where $z_0$,  $\alpha$ and $\beta$  set the depth,  low-redshift slope
and high-redshift cut-off for a given galaxy survey. We take $\alpha =
2$, $\beta = 2$ and $z_0 = 0.78$, yielding a median redshift of $z_m =
0.82$, similar to the CFHTLens Survey.

As the  particles in  our simulation are  in comoving  coordinates, we
transform this redshift distribution to a probability distribution for
the  particle  to  enter  our  catalogue  given  its  comoving  radial
distance,
 \begin{equation}
    p(r)  \propto  r^{\alpha}\left(\frac{dr}{dz}\right)  \exp\left[  -
      \left( \frac{z(r)}{z_0} \right)^{\beta}\right],
\label{eq:prob}
 \end{equation}
where
 \begin{equation}
    \frac{dr}{dz} = \frac{c}{H(z)},
 \end{equation}
and
\begin{equation}
  H(z)   =   \frac{H_0}{\left[\Omega_m(1+z)^3   +  \Omega_K(1+z)^2   +
      \Omega_{\Lambda}\right]^{1/2}},
\end{equation}
where $H_0$  is the  current Hubble value,  $\Omega_m$ is  the current
matter density,  $\Omega_\Lambda$ is  the current dark  energy density
and $\Omega_K$ is the curvature parameter.  Throughout we have assumed
a  flat, $\Omega_K=0$, cosmology  for our  simulations. We  sample the
particle  distribution so  our final  galaxy catalogue  has  a surface
density  of around  15  galaxies  per square  arcmin,  with a  maximum
redshift cut-off at $z=1.5$.

The left  panel of Figure \ref{fig:dndz}  is an example  of a redshift
distribution  taken from the  full particle  lightcone.  The  red line
shows  the  theoretical   distribution  from  equation  \ref{eq:prob},
normalised to  the number of  particles selected, that  the simulation
particles were drawn  from.  The clustered nature of  the particles in
the  distribution is  apparent as  the  peaks and  troughs around  the
theoretical curve can be seen.

\begin{figure}
 \psfig{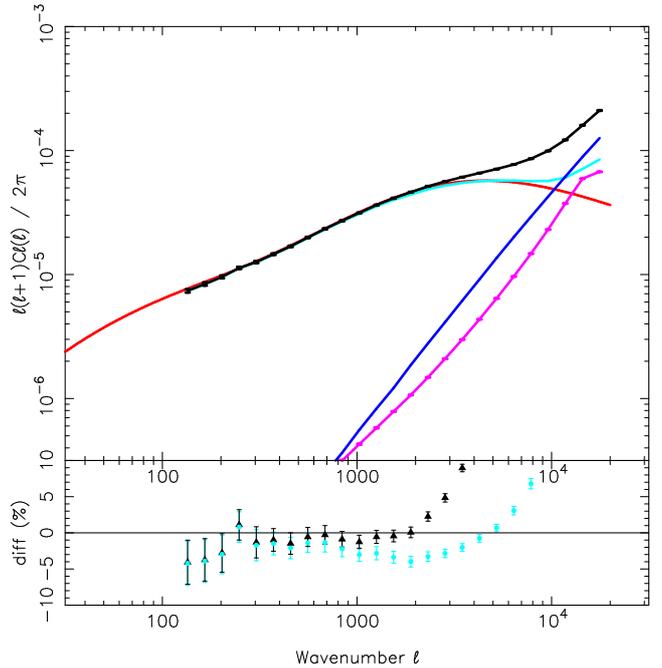}
 \caption{2D  shear power spectrum  for the  lightcone suite  with the
   $n(z)$ particle distribution. In  the upper panel, the smooth (red)
   line is the  theory prediction and the diagonal  (blue) line is the
   shot noise  prediction.  The  (black) points and  line is  the mean
   measured power spectrum  for the suite of mock  catalogues with the
   errors representing the error on the mean and the curve between the
   theory  prediction   and  the  measured  simulation   data  is  the
   shot-noise subtracted power  spectrum.  The diagonal (magenta) line
   shows the mean of the B-modes for the suite of mock catalogues with
   errors  on  the  mean.   The  bottom  panel  shows  the  percentage
   difference of  the data  from the theory  curve with errors  on the
   mean  (black)  and the  lower  (light  blue)  points represent  the
   shot-noise subtracted data.}
\label{fig:nz_PS}
\end{figure}

Our 3-D weak lensing catalogue  currently assumes that the redshift to
each  galaxy is  accurately known.  This  would be  appropriate for  a
spectroscopic  redshift survey,  but with  such large  surveys  we can
expect most weak lensing  catalogues will contain photometric redshift
estimates for each galaxy. To account for photometric redshift errors,
we randomly sample the measured  redshift from the true redshift using
a Gaussian distribution with uncertainty
\begin{eqnarray}
\sigma_z = \sigma_0(z_g)(1+z_g),
\end{eqnarray}
where $z_g$ is the true redshift  of the particle. For the purposes of
this work we assume a fixed $\sigma_0 = 0.05$. The right-hand panel of
Figure \ref{fig:dndz}  shows what the  distribution on the  left looks
like with  photometric redshift  errors.  The structures  are smoothed
out and the distribution becomes featureless. The photometric redshift
errors were implemented by specifying a Gaussian error.

Figure  \ref{fig:nz_PS} shows  the ensemble-averaged  2-D  shear power
spectrum estimated from 100 mock weak lensing surveys in the top panel
(black  dots)  with  errors  on  the mean,  compared  the  theoretical
prediction  in   red,  and  the  ensemble-averaged   B-mode  power  in
magenta. The (blue) diagonal  line shows the shot-noise prediction for
these  galaxy  redshift  distributed  lightcones. The  shot-noise  was
determined by  running the {\small  SUNGLASS} analysis on a  number of
simulation box volumes filled with randomly distributed particles. The
power spectrum of these  lightcones represents shot-noise estimate for
the simulations  and is a  remarkably straight power law.   The (light
blue) curve between the shot  noise and the measured power spectrum is
the shot-noise  subtracted power spectrum. The bottom  panel shows the
fractional difference between the average  of the mock surveys and the
theory curve,  with the error on  the mean (black)  and the shot-noise
subtracted points below  (light blue).  This shows that  the mock weak
lensing   survey  agrees   with  the   theoretical   expectation  from
wavenumbers  from $\ell=200$  to $\ell=2000$,  where  the disagreement
with theory can  be ascribed to the uncertainty  on the theory curve,
and the rise of shot-noise. The shot-noise subtraction in this case is
a few percent lower than the theoretical prediction and the reason for
this  is   not  well  understood   and  is  the  subject   of  ongoing
investigation.  The  analyses in  this  paper  will  use the  measured
simulation power spectrum only.  The  B-mode power appears to follow a
shot-noise  profile which is  consistent with  the effect  of sampling
from the full  particle lightcone.  A secondary source  for B-modes is
source clustering, which appears to be sub-dominant.

We found a dependence for the recovered shear and convergence power on
the number of pixels used to  estimate the 2-D lensing power. With too
many bins,  there were a number  of empty pixels and  this reduced the
amplitude of the  power spectrum. The amplitude of  the power spectrum
increased  with  fewer  empty  bins  before  converging  at  the  true
amplitude.  However, by using too few bins, the number of $\ell$ modes
recovered was reduced  due to pixelization effects. It  was found that
for this work, $768^2$ bins  provided a stable amplitude for the power
spectrum  with the largest  number of  modes possible  without causing
this amplitude  to fall. In this  case, 0.03\% of the  bins are empty.
If this number  is increased to 5\% empty, the  amplitude of the power
spectrum drops by  up to 10\%. This effect will  also be important for
observational  studies and  should be  considered when  binning survey
data to determine 2D lensing power spectra.

\section{Parameter Estimation}
\label{sec:Like}

\begin{figure*}
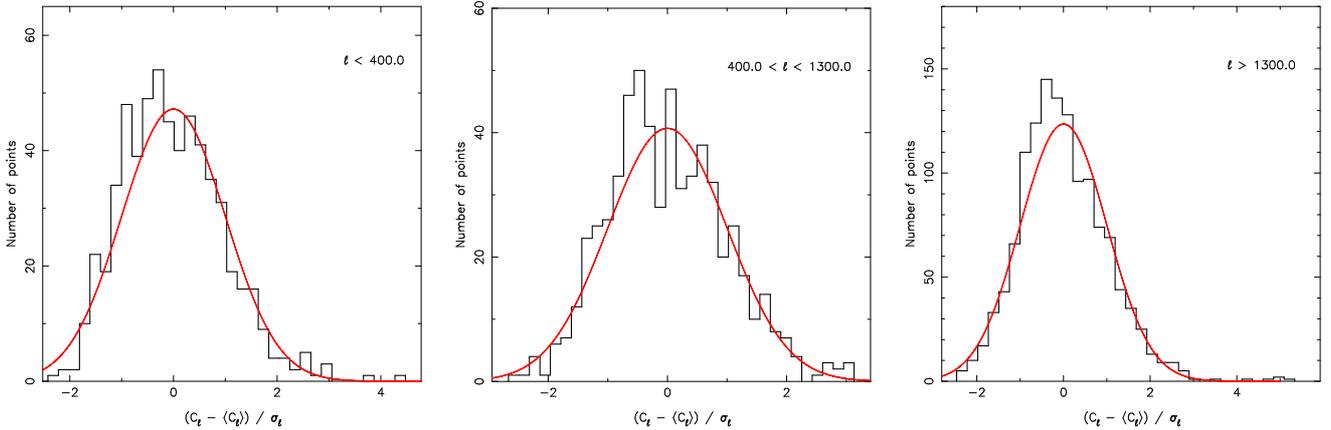

\centering
\begin{tabular}{ccc}
  \psfig{file=nz_Hist1_PS_gamma.ps,angle=-90,width=0.31\linewidth,clip=}
 &
  \psfig{file=nz_Hist2_PS_gamma.ps,angle=-90,width=0.31\linewidth,clip=}
 &
  \psfig{file=nz_Hist3_PS_gamma.ps,angle=-90,width=0.31\linewidth,clip=}
\end{tabular}
 \caption{Histogram of the distribution of power spectra for the suite
   of  lightcones with  the  $n(z)$ particle  distribution.  The  left
   panel  shows the  distribution of  the  $C_{\ell}^{\gamma \gamma}s$
   less   than   $\ell   =   400$,   the  middle   panel   shows   the
   $C_{\ell}^{\gamma \gamma}$  distribution from  $400 < \ell  < 1300$
   and the right panel shows the distribution from at $\ell > 1300$.}
\label{fig:nz_Hist}
\end{figure*}

As  described  in  the  previous  Section 100  simulations  have  been
generated  using  the  {\small  SUNGLASS} pipeline.  The  mock  survey
parameters are given in Table \ref{surveytable}.

For each of these mock lensing surveys the shear and convergence power
spectra has  been estimated,  and the ensemble  average power  and its
scatter  measured. Here  we want  to use  the mock  surveys to  test a
maximum   likelihood  cosmological   parameter   estimation  analysis,
typically used  to extract parameters from weak  lensing surveys. Here
we try and recover the amplitude of the matter clustering, $\sigma_8$,
and density parameter, $\Omega_m$, from a 2-D weak lensing survey.

In Section \ref{sec:mock} we showed that our simulations could produce
unbiased estimates of the shear power  from a mock survey over a range
of $\ell$-modes from $200$ to $2000$. For parameter estimation we need
to  know  the conditional  probability  distribution  of shear  power,
$p(\hat{C}_\ell^{\gamma\gamma}|\sigma_8,\Omega_m)$, for the likelihood
function, where  we have fixed  all other parameter at  their fiducial
values. This  is usually assumed to be  Gaussian \citep[although, see]
[who  study  non-Gaussian  likelihoods]{hss+09}.   Here we  test  this
assumption on our mock catalogues.  Figure \ref{fig:nz_Hist} shows the
distribution of  variations about the mean  of the $\hat{C}_{\ell}$'s,
$\Delta \hat{C}_\ell^{\gamma\gamma}$, divided by the ensemble-averaged
scatter in  the power, $\sigma(\hat{C}_\ell^{\gamma\gamma})$.   If the
distribution is  Gaussian, these distributions  should all lie  on the
unit-variance  Gaussian.  The  left  panel shows  a  histogram of  the
distribution of points for modes of $\ell < 400$ which is close to the
linear  region of  the  power  spectrum. The  middle  panel shows  the
distribution of  $C_{\ell}^{\gamma\gamma}$ for modes of $400  < \ell <
1300$ which  represents the non-linear  region of the  power spectrum.
The final panel  shows the distribution for modes  $\ell > 1300$ which
is the shot-noise  dominated regime. The smooth (red)  line in each of
the panels is a normalised unit-Gaussian curve. In each of the panels,
the histogram of points is peaked slightly to the left of the Gaussian
peak which  indicates a slight non-Gaussianity of  the distribution of
points.  This slight non-Gaussianity  may bias the Gaussian likelihood
analysis but  the dominant effect is currently  the inaccurate fitting
of the matter power spectrum  by the \cite{spj+03} formula at high $k$
\citep{gbs+10}.

The cosmological  parameters of  the simulations were  estimated using
Gaussian likelihood analysis where the likelihood is given by
 \begin{eqnarray}
  L(\hat{C}_\ell^{\gamma\gamma} |  \sigma_8, \Omega_m )  = \frac{1}{(2
    \pi)   ^{N/2}(\mathrm{det}~M_{\ell   \ell'})^{1/2}}  \exp   \left[
    \frac{-\chi^2}{2}\right],
 \end{eqnarray}
where
\begin{eqnarray}
  \chi^2   =  \sum_{\ell  \ell'}(\hat{C}_\ell^{\gamma\gamma}   -  \lgl
  C_\ell^{\gamma\gamma}                  \rgl)                 M_{\ell
    \ell'}^{-1}(\hat{C}_{\ell'}^{\gamma\gamma}          -         \lgl
  C_{\ell'}^{\gamma\gamma} \rgl),
\end{eqnarray}
and  $M_{\ell \ell'}$  is the  covariance  matrix of  the shear  power
spectra given by
\begin{eqnarray}
  M_{\ell   \ell'}   =   \lgl  \Delta   C_\ell^{\gamma\gamma}   \Delta
  C_{\ell'}^{\gamma\gamma} \rgl.
\label{eq:cov}
\end{eqnarray}
The inverse covariance matrix  was determined by performing a singular
value  decomposition (SVD)  on the  covariance  matrix \citep{ptv+92}.
The  resulting inverse covariance  matrix is,  however, biased  due to
noise in the covariance  matrix. \cite{hss07} propose a correction for
this bias by multiplying the inverse covariance matrix by a factor:
\begin{eqnarray}
\hat{M}^{-1}_{\ell \ell'} = \frac{N_S  - N_p - 2}{N_S- 1} M^{-1}_{\ell
  \ell'},
\end{eqnarray}
where  $N_S$  is the  number  of  simulations  used to  determine  the
covariance matrix, $N_p$  is the number of bins  in the power spectrum
and $\hat{M}^{-1}_{\ell \ell'}$ is the unbiased covariance matrix.

The  likelihood   analysis  relies  on  accurate   estimation  of  the
covariance matrix to show the degree of correlations.  The correlation
coefficients are 
 \begin{eqnarray}
  r_{\ell \ell'}  = \frac{M_{\ell \ell'}}{\sqrt{M_{\ell \ell} M_{\ell' \ell'}}}.
  \label{eq:ccm}
 \end{eqnarray}
The correlation  coefficient matrix is  equal to 1 along  the diagonal
and the  off diagonal components  will show how correlated  the $\ell$
modes are, with  numbers close to zero indicating  low correlation and
numbers close to (minus) one indicating high (anti-)correlation.

\begin{figure}
  \psfig{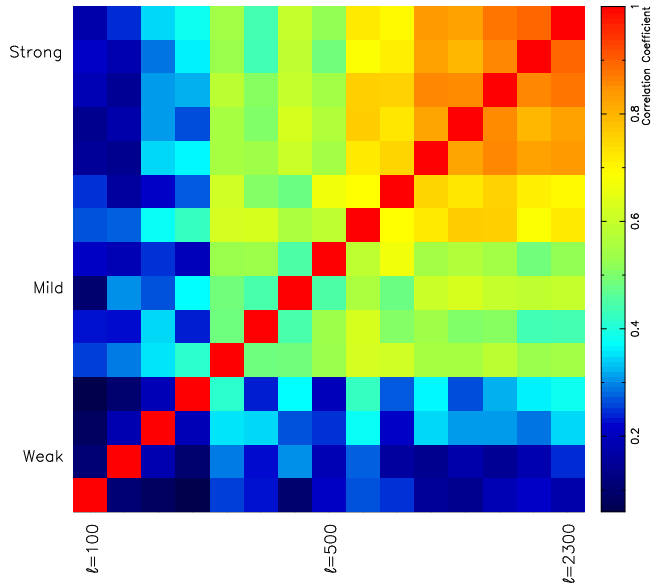}
  \caption{Correlation  coefficient  matrix.   This figure  shows  the
    correlation between  the bandpower $\ell$-modes  in the covariance
    matrix.  The  higher  $\ell$  bandpowers are  strongly  correlated
    (shown  in  reds), while  the  lower  bandpowers  are only  weakly
    correlated (shown in blues).}
  \label{fig:CCM}
\end{figure}

Figure \ref{fig:CCM} shows the  correlation coefficient matrix for the
$\ell$ modes being considered between  $100 < \ell < 2500$.  The modes
with a low correlation are represented in black and dark blues and the
modes with  a high correlation shown  in yellows and  reds. This shows
the the bandpowers at low  $\ell$ have very little correlation between
them, as we would expect, since for an all-sky survey the linear power
is  uncorrelated. At higher  $\ell$ bandpower,  the modes  become more
correlated,  due  to  cross-talk   between  different  scales  due  to
nonlinear clustering in the  matter power spectrum.  The variations in
this  coefficient matrix  indicate an  error of  around 10\%  which is
suitable for the  studies in this paper. This error  can be reduced by
introducing more  realisations into the calculations.  In our analysis
we shall  consider modes  up to $\ell  = 1500$, where  the correlation
coefficient is around $r_{\ell \ell'} \approx 0.6$ .

\begin{figure}
  \psfig{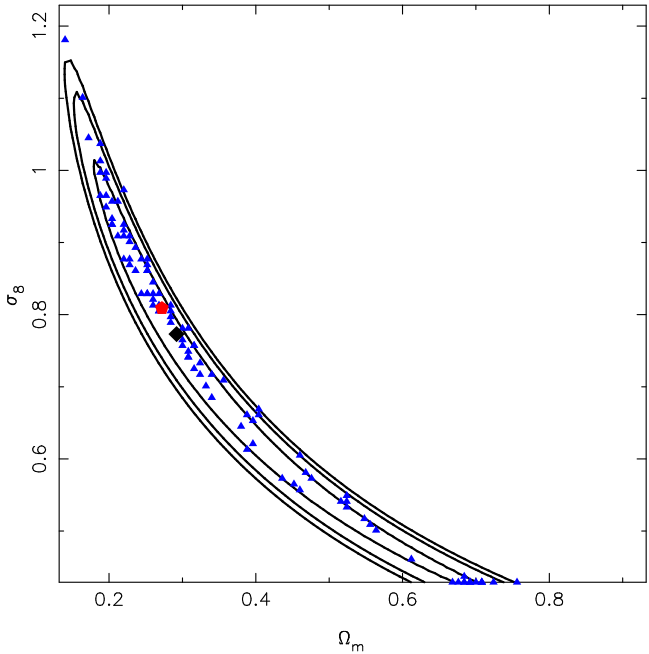}
  \caption{Gaussian likelihood estimate.  The black contours come from
    the  simulation with  the  closest fit  to  the true  cosmological
    parameters.   The blue  triangles show  the best  fit cosmological
    parameters  for the  suite of  lightcones.  The  true cosmological
    parameters are shown at the  red polygon and the combined $\chi^2$
    best fit parameter is shown at the black diamond. }
  \label{fig:Best}
\end{figure}

Figure   \ref{fig:Best}  shows   the   $\chi^2$-distribution  in   the
$\sigma_8$-$\Omega_m$ plane for our ensemble of simulations. The black
lines represent the $\chi^2$  two-parameter, 1, 2 and 3$\sigma$ (which
should  contain 68.3\%,  95.4\% and  99.7\% of  the points  assuming a
bivariate Gaussian distribution), contours  of parameter space for the
cosmological  parameters. However,  this  clearly is  not a  bivariate
Gaussian distribution. The contours  shown are representative and come
from the simulation that had the best fit parameters that were closest
to the  true input  parameters (the point  shown by the  red polygon).
The blue triangles  represent the best fit points for  each of the 100
realisations.  With  this distribution, 68\% of the  points lie within
the  1$\sigma$ contour,  93\% within  the 2$\sigma$  contour  and 97\%
within the  3$\sigma$. The black  diamond represents the best  fit for
the combined $\chi^2$ estimate as discussed below.

\begin{figure}
  \psfig{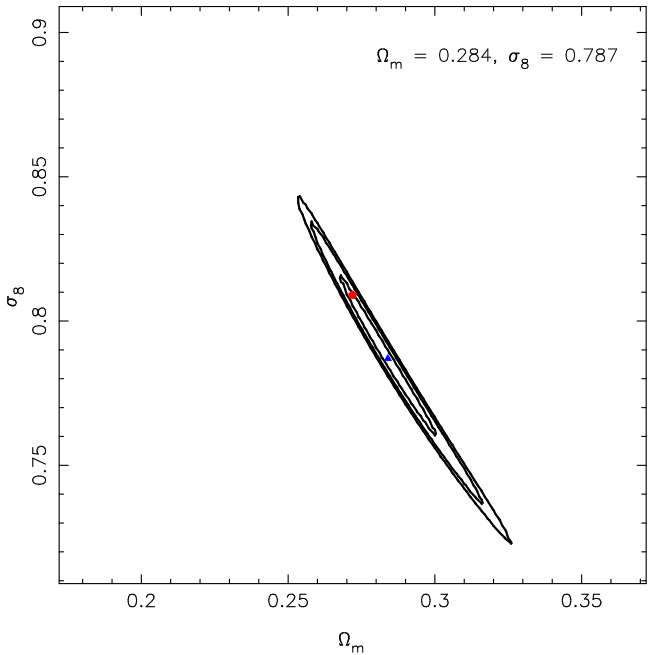}
  \caption{Combined  $\chi^2$  likelihood. The  black  lines show  the
    combined $\chi^2$ 1, 2  and 3$\sigma$ contours.  The blue triangle
    shows the  best fit parameters  for the combined $\chi^2$  and the
    red star shows the true cosmological parameters.}
  \label{fig:Total}
\end{figure}

The results  from this analysis  give us very encouraging  results for
the parameter estimation. Figure  \ref{fig:Total} shows the results of
combining the likelihoods for all  100 realisations, as if we have one
hundred independent 100  square degree surveys. Even for  this test we
see the  maximum likelihood recovered parameter values  lie within the
$1-\sigma$ confidence contour. The  marginalised error on the measured
parameters for the combined 100 surveys is $\Delta \Omega_m=0.012$ and
$\Delta   \sigma_8=0.022$,  within  expected   errors.  There   is  no
significant bias in this result at the level of $\sim 0.02$.

\section{Discussion and Conclusions}
\label{sec:disc}
This work introduces  {\small{SUNGLASS}} -- {\small S}imulated {\small
  UN}iverses  for  {\small  G}ravitational  {\small  L}ensing  {\small
  A}nalysis and {\small S}hear  {\small S}urveys. {\small SUNGLASS} is
a new,  rapid pipeline that generates  cosmological N-body simulations
with  {\small  GADGET2}. It  computes  weak  lensing  effects along  a
lightcone using line-of-sight integrations  with no radial binning and
the  Born approximation  to  determine the  convergence  and shear  at
multiple source redshifts. This information is interpolated back on to
the particles in the lightcone to generate mock shear catalogues in 3D
for testing weak lensing observational analysis techniques.

In this work,  {\small SUNGLASS} was used to  generate 100 simulations
with $512^3$ particles,  a box length of $512  h^{-1}$~Mpc and a WMAP7
concordance  cosmology. The corresponding  mock shear  catalogues were
100  sq  degrees  with  a  source redshift  distribution  with  median
$z_m=0.82$ and  15 galaxies per square arcminute.   The parameters are
easily changed within the {\small  SUNGLASS} pipeline so that the mock
shear catalogues matches the survey of interest.

To  show the  reliability  of the  lightcones  generated with  {\small
  SUNGLASS}, E- and B-mode power spectra were shown at multiple source
redshifts. The results show that  at low redshifts, the signal becomes
dominated  by shot-noise  at reasonably  low $\ell$.   With increasing
source   redshift,  the  power   spectrum  recovers   the  theoretical
prediction over a wider range of modes, $\ell < 2500$.

Given that the  measured power spectrum of the  simulations appears to
follow the  predicted shot noise at  higher modes, the  shot noise was
subtracted from the power spectra to increase the recovered range. The
theoretical prediction is expected to under predict the power spectrum
around  the  turn over  and  consequently,  the  simulations could  be
recovering the power  spectrum up to around $\ell =  5 \times 10^4$ at
the highest redshift planes.

The  multiple   source  redshift  plane  shear   and  convergence  was
interpolated onto  the particles in  the lightcone to generate  a mock
shear  catalogue.   A  redshift  sampling  was  also  imposed  on  the
lightcone  to   mimic  an  observed  shear   catalogue.  Binning  this
distribution too finely resulted in empty bins which had the effect of
suppressing the power spectrum.  This has implications for observations
where the number of objects  per square arcminute should be taken into
account, as well  as the density of the  binning, when determining the
accuracy of the power spectrum.

The mock shear catalogues were used to determine a covariance matrix
which is essential for both parameter estimation and data analysis. A
strength of {\small SUNGLASS} is the ability to rapidly produce Monte
Carlo realisations of these catalogues, ensuring independent mock data
sets for the generation of the covariance matrices.

The  mock catalogues  were also  used  to perform  a simple  parameter
estimation  using Gaussian likelihood  analysis.  The  distribution of
power spectra were  shown to be reasonably Gaussian  and the resulting
parameter estimation  contours for a single realisation  showed a good
agreement  with the input  parameters within  the 2-parameter  1,2 and
3$\sigma$ error contours.

The  combined  likelihood  from   the  100  simulations  shows  narrow
likelihood  contours  and   accurate  parameter  recovery  within  the
expected errors, with no evidence of significant bias at the level of
$\sim 0.02$.

Current and  future telescope surveys  promise to provide  an enormous
amount  of data for  weak lensing  analysis. Weak  lensing is  still a
young field and  analysis techniques are still being  developed. It is
essential that  the strengths and  weaknesses of these  techniques are
fully  understood  before  using   them  on  real  data  with  unknown
parameters.   Using   the  simulations,  lightcones   and  mock  shear
catalogues   provided   by  the   {\small   SUNGLASS}  pipeline,   and
demonstrated  in  this  paper,  is  an excellent  way  to  test  these
observational weak  lensing analysis  techniques. The outputs  of this
pipeline have  been rigorously tested and are  well understood, making
them  ideal for generating  covariance matrices  that are  critical to
many observational analysis techniques.

\section*{Acknowledgments}

AK  acknowledges  the  support  of  the  European  DUEL  RTN,  project
MRTN-CT-2006-036133.  AK  would also  like to thank  Benjamin Joachimi
for the use of his power spectrum theory code as well as John Peacock,
Richard  Massey, Tom  Kitching  and Martin  Kilbinger  for their  very
helpful discussions on this work.

\bibliographystyle{mn2e}
\bibliography{thesis}

\appendix

\section{Derivation of the line-of-sight convergence with no radial binning}
\label{ap:conv}
This  appendix  shows  how  the  line-of-sight  convergence  shown  in
equation \ref{eq:convergence} was derived.

Start with the general equation for the convergence,
\begin{equation}
  \kappa = \int_0^{r_s}dr~K(r,r_s)~\delta(\rb),
  \label{eq:kap}
\end{equation}
where  $r_s$ is  the  lensing source  redshift,  $\delta(\rb)$ is  the
fractional matter overdensity and $K(r,r_s)$ is the kernel
\begin{equation}
K(r,r_s) = \frac{(r_s-r)r}{r_sa(r)}\frac{3H_0^2\Omega_m}{2c^2}.
\end{equation}
The overdensity $\delta(\rb)$ is given by
\begin{equation}
\delta(\rb) = \frac{n(\rb)}{\bar{n}(r)} - 1,
\end{equation}
where  $\bar{n}(r)$ is  the  average density  at  the comoving  radial
distance $r$ and is constant in comoving co-ordinates.\\

The particle number  density, $n(\rb)$, is given by a  sum of 3D delta
functions
 \begin{equation}
  n(\rb)    =     \sum_{i=part}\delta^{3D}(\rb-\rb_i)    =    \sum_{i}
  \frac{\delta^{1D} (r-r_i)} {r^2} \delta^{2D} (\thetab - \thetab_i),
 \end{equation}
where  $part$ are  the  particles in  the  pixel with  $r_i \le  r_s$.
Substituting this sum  of delta-functions into equation (\ref{eq:kap})
yields  the average convergence  per pixel  on the  sky, $p$,  with no
radial binning;
\begin{equation}
\bar \kappa_{p} = \frac{1}{\Delta \Omega}\int_{p}\! d^2\theta \,\kappa
=    \sum_i    \frac{k(r_i,r_s)}{\Delta\Omega_p\bar{n}(r_i)r_i^2}    -
\int_0^{r_s} dr~k(r,r_s),
\end{equation}
where $\Delta \Omega_p = \Delta \theta_x \Delta \theta_y$.

\end{document}